# Direct observation of the morphological changes in a carbon fiber composite by means of *in-situ* micro-CT compression experiment

M. I. Santos[1,2], K. Koschek[3], L. Pursche[3,] J. M. de Souza e Silva[2]

[1] Institute of Physics, Martin-Luther-Universität Halle-Wittenberg, Halle, Germany

[2] Fraunhofer Institute for Microstructure of Materials and Systems IMWS, Halle (Saale), Germany

[3] Fraunhofer Institute for Manufacturing Technology and Advanced Materials IFAM, Bremen, Germany,

**Abstract**

Carbon fiber reinforced polymers are widely used due to their lightweight and strong structural properties and understanding their mechanical behavior is crucial for their reliable use in structural components. In this study, we evaluated the mechanical properties of a carbon fiber reinforced composite made of a dynamic polymer based on benzoxazine and polyetheramine Jeffamine® ED 600. Standard compression tests were performed on specimens with layers oriented longitudinally or transversely to the loading direction. In-situ mechanical testing using micro-computed tomography (micro-CT) imaging was conducted to capture detailed images of the internal microstructure during deformation. Changes in porosity and pore shape were observed as the composite underwent compression. Digital Volume Correlation (DVC) analysis was applied to quantify displacements and strains within the material and enabled the visualization and quantification of strain patterns, which agree with kinking failure.

**Introduction**

Polymer matrix composites are widely utilized in applications that require lightweight and strong structural materials. These composites offer a combination of high stiffness and strength, making them highly desirable for transportation purposes on land, in the air, and on water. [1] In particular, carbon fiber reinforced polymer composites have attracted attention in various industries for their exceptional mechanical properties, including high specific stiffness and strength. [2, 3] A thorough investigation of carbon reinforced polymer composites' mechanical performance is critical to understanding their behavior and ensuring their reliable use in structural components. The aim is to achieve a detailed and robust understanding of the deformation, damage, and fracture behavior under different loading conditions and rates.

Various methods are used to assess these properties, such as tensile, flexural and impact tests. [1] These tests provide valuable information about the strength, stiffness, toughness, and durability of the composite under different loading conditions. [4, 5] In addition, strain measurements done using digital images can be used to generate high-resolution surface maps of the displacement and strain field, known as Digital Image Correlation (DIC), to gain insight into the microstructural effects on the materials due to an applied strain. [1] As an advancement of the now well-established and standardized DIC, Digital Volume Correlation (DVC) extends this methodology to three-dimensions by using sequential volumetric images recorded *in-situ* as inputs to deliver three-dimensional displacement and strain fields within the structure. [6]

*In-situ* mechanical testing using computed tomography (CT) imaging has emerged as a powerful tool for investigating the behavior of carbon-reinforced polymer composites under load. [7, 8] By combining mechanical testing devices with CT scanners, it is possible to capture detailed images of the internal microstructure of the composite during deformation. [9, 10] DVC is then used to analyze the deformation and displacement of materials in three-dimensions (3D). It involves comparing digital volumes datasets of the material under analysis before and after its deformation to calculate the correlation between the reference (before deformation) and deformed images, and by tracking the correlation, it can quantify displacements and strains within the material. [11] This technique provides a better understanding of the mechanical response of the composite, including localized deformation, crack propagation and failure mechanisms.

In reinforced polymer composites, the interface between the polymer matrix and the reinforcing fiber plays a crucial role in short- and long-term performance [12]. Interfacial regions are often prone to failure due to the interaction of different damage mechanisms that occur in three-dimensional space [13] [14] and are, thus, critical for optimizing their mechanical performance.

Therefore, in this study, we evaluated the mechanical properties of a carbon fiber reinforced composite made of a dynamic polymer based on benzoxazine and polyetheramine Jeffamine® ED 600, [15] fabricated in individual layers as prepregs that were consolidated into a single composite with a thickness of 15 mm. The mechanical properties of this material were evaluated using standard testing methods on a universal testing machine by compression tests on specimens with layers oriented longitudinally or transversely with respect to the loading direction (z-direction). Also, we conducted *in-situ* mechanical testing with micro-computed tomography (micro-CT) imaging of one composite specimen with transversely oriented carbon fibers. Our results show details in the changes in porosity of the composite as it underwent compression and revealed the occurrence of changes in the shape of the pores. We utilized DVC [16] to visualize and quantify strain patterns, which are essential for understanding the mechanical behavior and response of the material under load. Our findings provide insights on the performance and failure mechanisms of the polymer based on benzoxazine and polyetheramine, aiding in the design and optimization of this composite material.

**Experimental Section**

*Materials*

Araldite® MT 35600 (BA-a) was provided by Huntsman Advanced Material (Huntsman International LLC, Texas, USA), Jeffamine® ED-600 were purchased from Merck KGaA (Darmstadt, Germany). Aluminum plate for the sample manufacturing were coated with the release agent ACMOScoat 82-9101 (ACMOS CHEMIE KG, Bremen, Germany). Carbon fibers (UD, 603 g/m$^2$) were obtained from Saertex (SAERTEX GmbH & Co. KG, Saerbeck, Germany)

*Sample preparation*

The carbon fiber reinforced composite was prepared in two steps. The first step, the single layer manufacturing, consisted in liquifying and homogenizing BA-a and Jeffamine® ED-600 at molar ratios of 1:0.25 (BA-a:Jeffamine®) at 120 °C in a convection oven (Heraeus 19 Function Line, UT 6 P/ UT 20 P, 250 °C, Thermo Fisher Scientific Inc., Waltham, USA). The monomer mixture was degassed for 30 min in a vacuum oven (Heraeus Vacutherm VT6025, Thermo Fisher Scientific Inc., Waltham, USA). Single fabric plies with the dimensions 120 mm × 120 mm were laid on a coated aluminum plate and sealed in a vacuum bag setup. Infusion was performed in a convection oven (Vötsch VTL 100/150, Vötsch Industrietechnik GmbH, Reiskirchen, Germany) at a temperature of 120 °C. The resin polymerized under the applied vacuum at 120 °C and 150 °C, for 2 h at each temperature. In a second step, the composite was fabricated to a thickness of 15 mm. For this, 27 individual layers prepared in the previous step were stacked and consolidated in a heating press (Wickert WLP 200 S, WICKERT Maschienenbau GmbH, Landau, Germany) at 150 °C with a pressing force of 20 kN for 40 min. Specimens were cut using a wet cut-off sander (Conrad Apparatebau GmbH, Clausthal-Zellerfeld, Germany) and were wet ground into a cylindrical shape (180 grit) having the layers either vertically or horizontally oriented.

For the preparation of the unreinforced reference specimen, cylindrical silicone molds with a diameter of 15 mm were used to produce pristine polymer samples. BA-a and Jeffamine® ED-600 were mixed in a molar ratio of 1:0.25 and homogenized at 120°C in a convection oven. After 30 minutes of degassing in a vacuum oven, the liquid resin was poured into the silicone mold. After curing at 120 °C for 2 h and 150 °C for 4 h, the approximately 15 mm long, cylindrically shaped rod was demolded. The specimens were cut into 15 mm high samples using a wet-cutting grinder.

*Uniaxial compression tests*

To determine the mechanical properties of the composite, uniaxial compression tests were carried out on cylindrical specimens with the dimensions given in Table 1. The test was carried out in accordance with the DIN-EN-ISO 604 standard, [17] with n = 3 for each specimen, in an AllroundLine Z050 universal testing machine (Zwick Roell, Ulm), at 22.1°C and 61.1% relative air humidity.

**Table 1.** Dimensions of the samples used in the uniaxial (vertical) compression tests.

| Sample | Height - H (mm) | Diameter - D (mm) | H/D |
|---|---|---|---|
| Unreinforced polymer (pristine) | 15.17 ± 0.07 | 16.09 ± 0.06 | 0.94 ± 0.01 |
| Transverse fibers[a] | 13.20 ± 0.20 | 14.99 ± 0.02 | 0.89 ± 0.01 |
| Longitudinal fibers[b] | 14.76 ± 0.04 | 13.90 ± 0.40 | 1.07 ± 0.03 |

Orientation of the fibers related to compression axis (z): a) θ = 90°, b) θ = 0°

The specimens were subjected to a pre-load of 0.1 MPa. To obtain the compressive modulus, an initial strain rate of *v = 0.02 l* (mm/min) was applied, where *l* = length in mm. To test the strain rate, it was increased to *v = 0.5 l* (mm/min). The Young's modulus ($E_c$) was measured between 0.25% and 0.5%. The nominal stress ($\sigma_B$) and the compressive strain at break ($\varepsilon_B$) were determined at the first observed fracture event. The work to produce the first fracture in the specimens was obtained by integrating force versus displacement curves (not shown) in the region between $\varepsilon_0$ and $\varepsilon_B$. In addition, the fracture energy was estimated by dividing the work by the fracture area, obtained here by multiplying the sample diameter by its length, as the cracks propagate throughout the specimen length. [18]

*Intermittent in-situ compression test in X-ray micro-CT*

Imaging experiments were performed in a RayScan 200E using a chromium X-ray source at 126 kV and 140 µA with absorption contrast. Four full tomography datasets were obtained, each with an exposure time of 999 ms per projection and a total of 1440 projections over 360°. With the sample placed in a Deben Microtest Tomography tension/compression stage, the first tomography was taken from the reference state without compressing the sample between a top, static jaw and a bottom, moving jaw. The compression was controlled by the axial displacement of the lower jaw of the stage. The second, third and fourth full tomographies were taken after compression increments.

*Image processing and Digital Volume Correlation*

Image reconstruction was performed by a filtered back-projection algorithm using the RayScan's built-in software. The final 16-bit images had a voxel size of 12.1 µm. Commercial software Avizo (Thermo Fisher Scientific, version 3D 2023.1) was used for image correction, segmentation, DVC analysis and 3D rendering. The contrast of the datasets was adjusted using the mean grey value of the dataset of the uncompressed sample as a reference. A sub-volume of $566 \times 550 \times 795$ voxels was extracted from each dataset to remove boundaries while retaining the largest volume available. Automatic image registration was performed in Avizo using the Image Registration Wizard. A feature of $50 \times 100 \times 60$ voxels (corresponding to a crossing of 2 fibers) was used as reference for registration and is located on the top of the specimen and is visible in all datasets (Fig. S1). The Avizo Digital Volume Correlation algorithm (global approach) was run with a cell size (sub-volume edge length) of 1500 voxels, (defined after running a study of the influence on the correlation coefficient as described in [19, 20]) and a convergence criterion of 0.001. Displacement vectors and strain maps were obtained. Further analysis included extraction of a smaller sub-volume containing only the sample and image processing to separate the solid phase from the porous phase by first applying a median filter (1 iteration), followed by iso-data threshold segmentation. After segmentation to separate pixels related to the sample from those related to the air, the sample volume fraction was estimated for each xy slice in each dataset and corresponds to the ratio of the number of pixels corresponding to the sample to the total number of pixels in the slice.

**Results**

Using standard uniaxial compression tests, we have evaluated the mechanical properties of the composite materials containing fibers oriented either longitudinally or transversely to the compression axis in comparison to the pristine material. When the fibers are longitudinally oriented, the composite behaves as a rigid polymer and does not crack during compression, even under an applied force of 47.5 kN (Fig. 1 a, Table 2). This material exhibits the highest modulus of compression of any of the materials tested in this study, 265% higher than the pristine polymer. On the other hand, the presence of carbon fibers oriented transversely to the compression axis reduces the compressive modulus by 53% compared to the pristine polymer. Furthermore, the first fracture point occurs at approximately 10% strain in both weaker materials (Fig. 1 a), however the work required for failure and the corresponding fracture energy are at lower levels when fibers are present and transversely oriented. The stress-strain curve of this material shows significant irregularities due to the progressive failure mechanism observed in carbon fiber composites. [21]   During the test, the material experiences the development of kink bands (Fib. 1 b, blue arrows) and there is an abrupt drop in the applied load. Inspection of the failed composite specimen reveals delamination due to shear stress. An SEM image of the delaminated surface (Fig. 1 c) shows the imprint of the fibers, indicating that the delamination occurs between compressed layers. The specimen containing carbon fibers oriented longitudinally to the compression appears unchanged after the compression experiment, while the pristine specimen deformed completely after compression (Fig. 1 b, top to bottom).

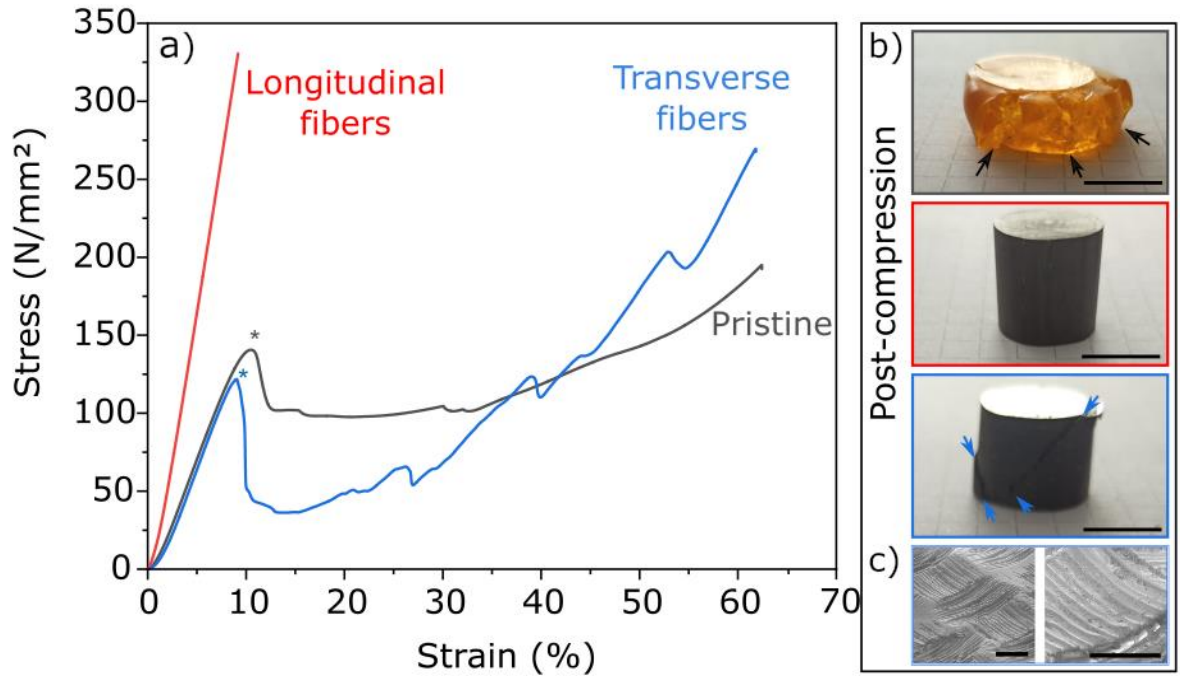

**Figure 1.** a) Representative stress-strain curves from uniaxial compression tests. First fracture events are highlighted with asterisks, b) photographs after compression to show the fractures, highlighted with arrows (scale bars: 1 cm), c) SEM images of the surface of the fracture in the sample with carbon fibers transverse to the compression axis (scale bars: 100 μm).

**Table 2**. Summary of uniaxial compression test results.

| Sample | $E_c$ (MPa) | $s_B$ (MPa) | $e_B$ (%) | Work to failure $W_f$ (J) | Fracture energy $G_f = W_f/A$ (N/mm²) |
|---|---|---|---|---|---|
| Pristine | 438 ± 161 | 135 ± 7 | 10.1 ± 0.4 | 22 ± 2 | 90 ± 8 |
| Transverse fibers | 206 ± 25 | 128 ± 9 | 9.5 ± 0.6 | 14 ± 3 | 70 ± 14 |
| Longitudinal fibers | 1163 ± 334 | - | - | - | - |

We decided to further analyze the sample containing transversely oriented fibers, due to its weaker nature (Table 2), with *in-situ* compression in a micro-CT to observe how it deforms under stress and possibly identify weaknesses and vulnerabilities in the material. We performed full 3D scans of the sample in a total of five times: step 0 is the initial state before any compression, and 1 to 4 are successive compression steps applied while controlling the sample displacement in the z-axis (Fig. 2).

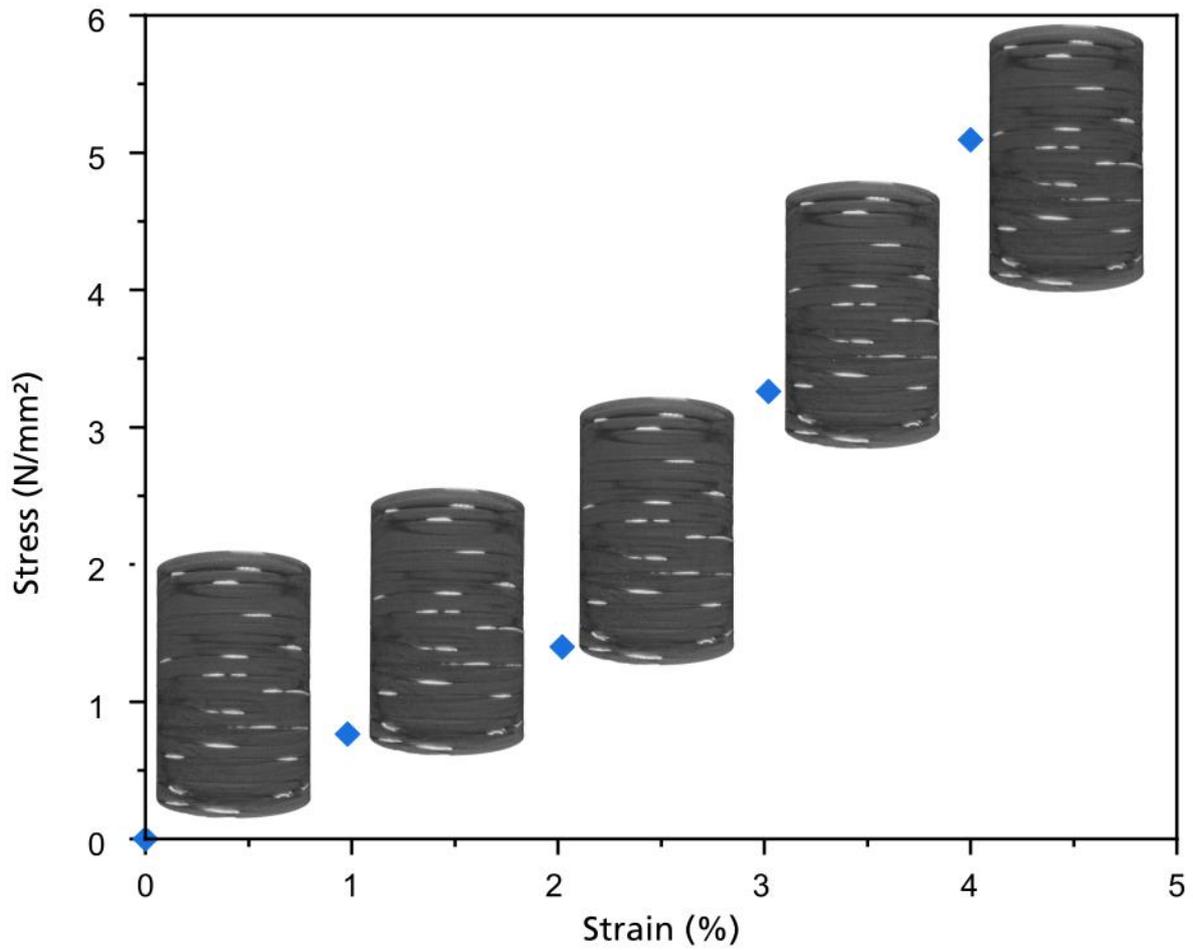

**Figure 2.** Stress-strain curve from uniaxial compression test performed *in-situ* in a micro-CT, of a sample containing fibers oriented transversely. Volumetric renderings of the specimen are shown in the points in which the scan was performed.

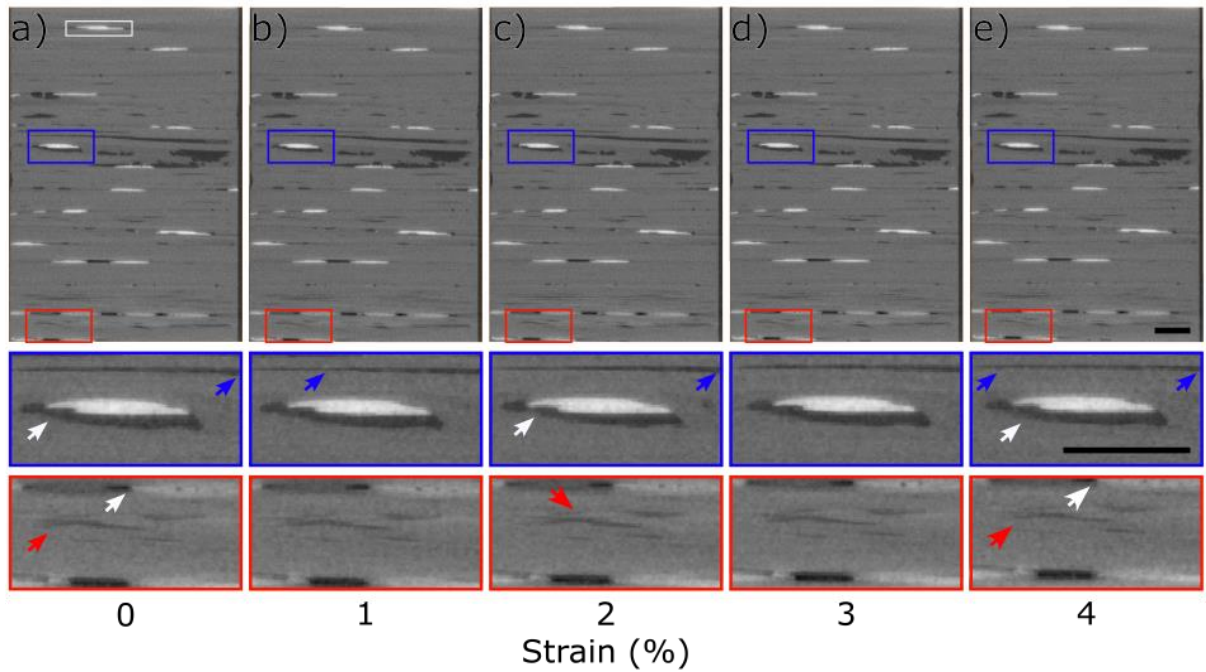

**Figure 3.** Frontal virtual slices in a middle section of the specimen submitted to *in-situ* compression in a micro-CT. a) Uncompressed, and b-e) after successive incremental compressions. Area within white rectangle in a) indicate the structure used as reference for registration of the datasets. Areas highlighted within blue and red rectangles are shown enlarged immediately below. Scale bars: 1000 µm.

Vertical virtual slices provide a detailed view of the specimen, showcasing pores, fibers, and matrix in distinct grey levels (Fig. 3). This visualization allows us to observe the changes that occur during compression. By tracking specific pores, we can observe a slight vertical displacement (ca. 30 µm – 60 µm between uncompressed and final states) and changes of shape (such as in Fig. 3, white arrows in red and blue rectangles). Additionally, the small spaces between plies reduce (Fig. 3, blue arrows). We can visualize the spaces between plies can resemble waves (Fig. 3, red arrows), undergo shape alterations, and may even disappear partially after the fourth compression. Segmentation of the data (Fig. 4) facilitates visualization of the larger pores, which change in shape and vertical position due to compression, and allows quantification of these changes in terms of change in volume fraction per xy slice (Fig. 5a).

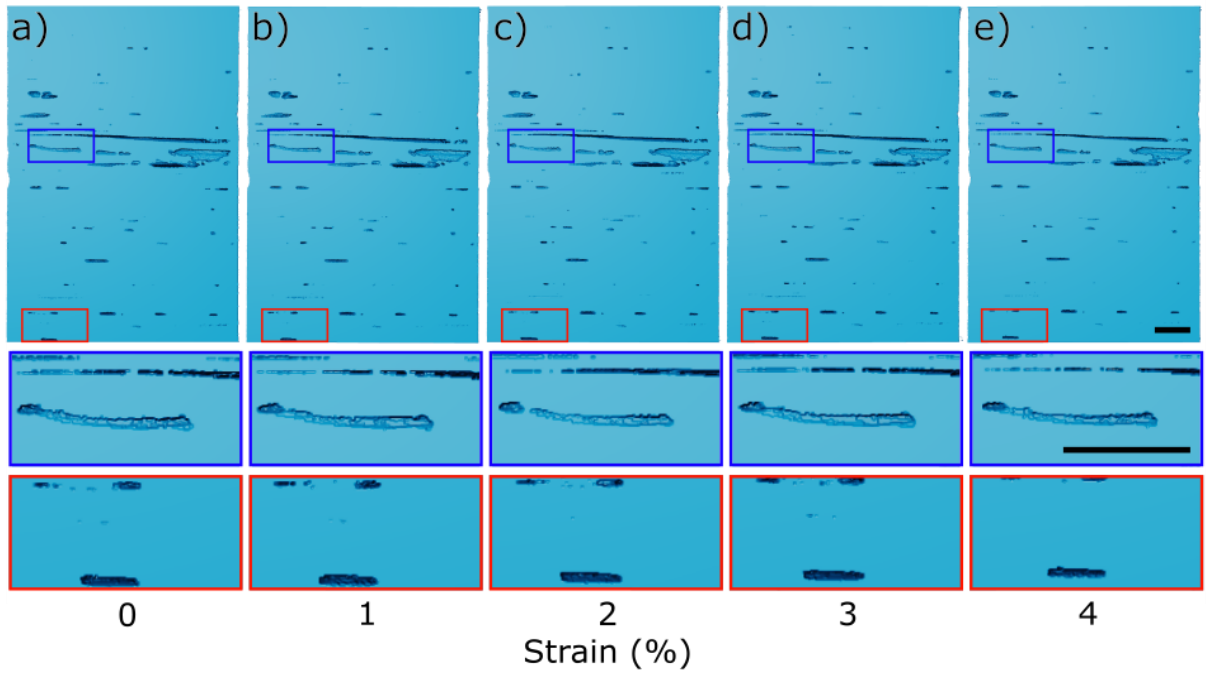

**Figure 4.** Volumetric rendering of the specimen (a) uncompressed and (b-e) after successive compression steps. The areas highlighted in the blue and red rectangles are shown in enlarged form immediately below. Scale bars: 1000 μm.

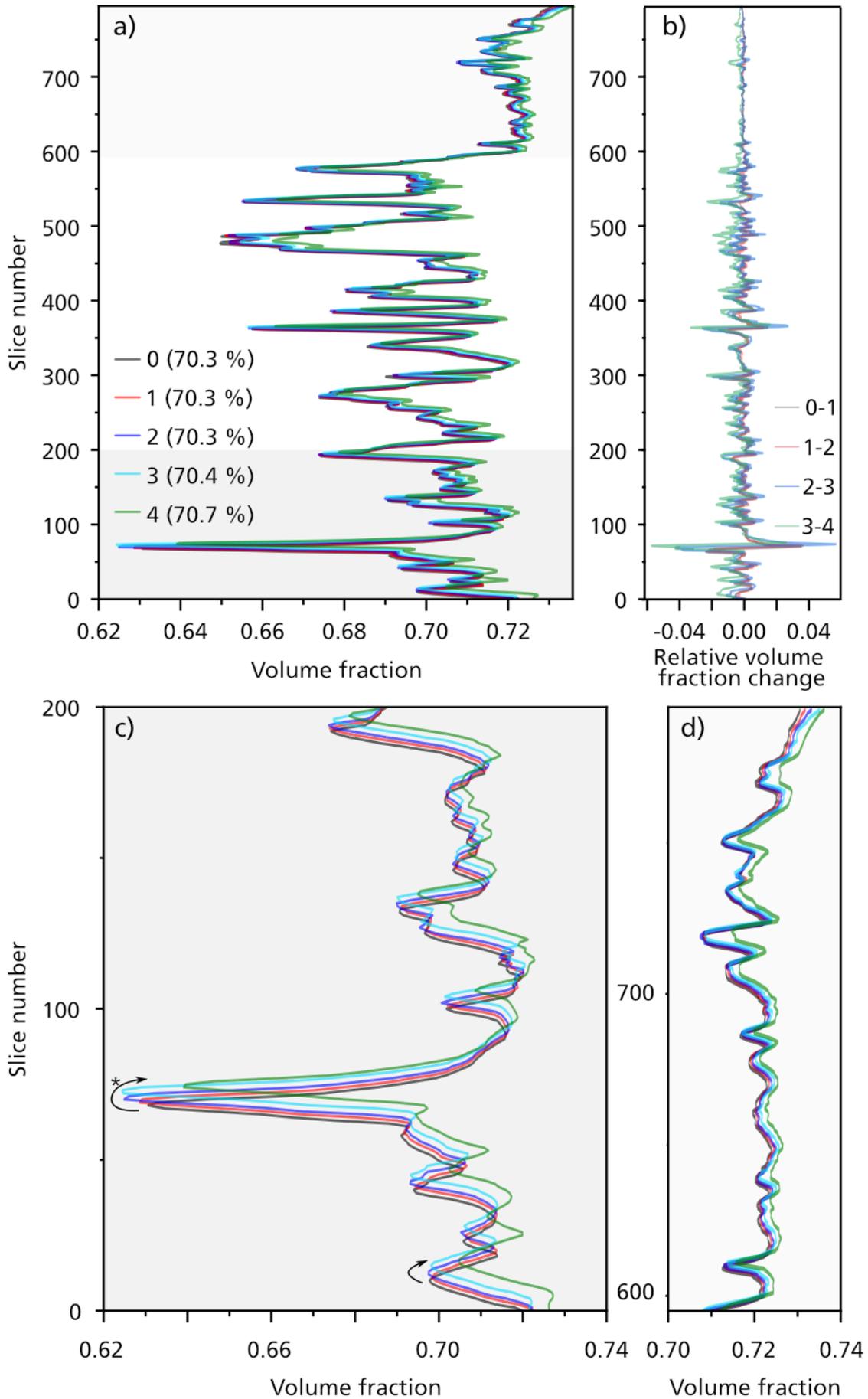

**Figure 5.** Sample volume fraction calculated for a) each slice orthogonal to the loading z-axis at different compression states (0: uncompressed, 1-4: consecutive compressions), b) relative change of volume fraction for corresponding slices between compression states (0-1, 1-2, 2-3, 3-4), c) volume fraction for the bottom 200 slices, and d) the 200 slices on the top.

A more detailed analysis of the relative change in volume fraction between successive states shows that the bottom of the specimen experiences a more pronounced effect of compression (Fig. 5 b, c). In addition, a progressive vertical displacement of the pores is observed (Fig. 5 c, arrows). This behavior is different in the upper and lower part of the specimen, with the lower part (red rectangles in Fig. 4, and Fig. 5 c) showing a more intense vertical displacement of the pores compared to the upper part of the specimen (blue rectangles in Fig. 4, and Fig. 5 d). A progressive decrease in volume fraction along the entire sample indicates reduction of pore volume due to compression, which is expected.

The inspection of the tomographic slices corresponding to the most prominent change observed in the volume fraction (Fig. 5 c, asterisk, and Fig. 6) shows the carbon fibers grouped together resembling tapes (Fig. 6 b-d). Under compression, subtle changes can be observed in both the tapes, as the attachment and separation of fibers appeared to be a dynamic process, with the positions of separated fibers changing between consecutive tomograms (Fig. 6 c, arrows).

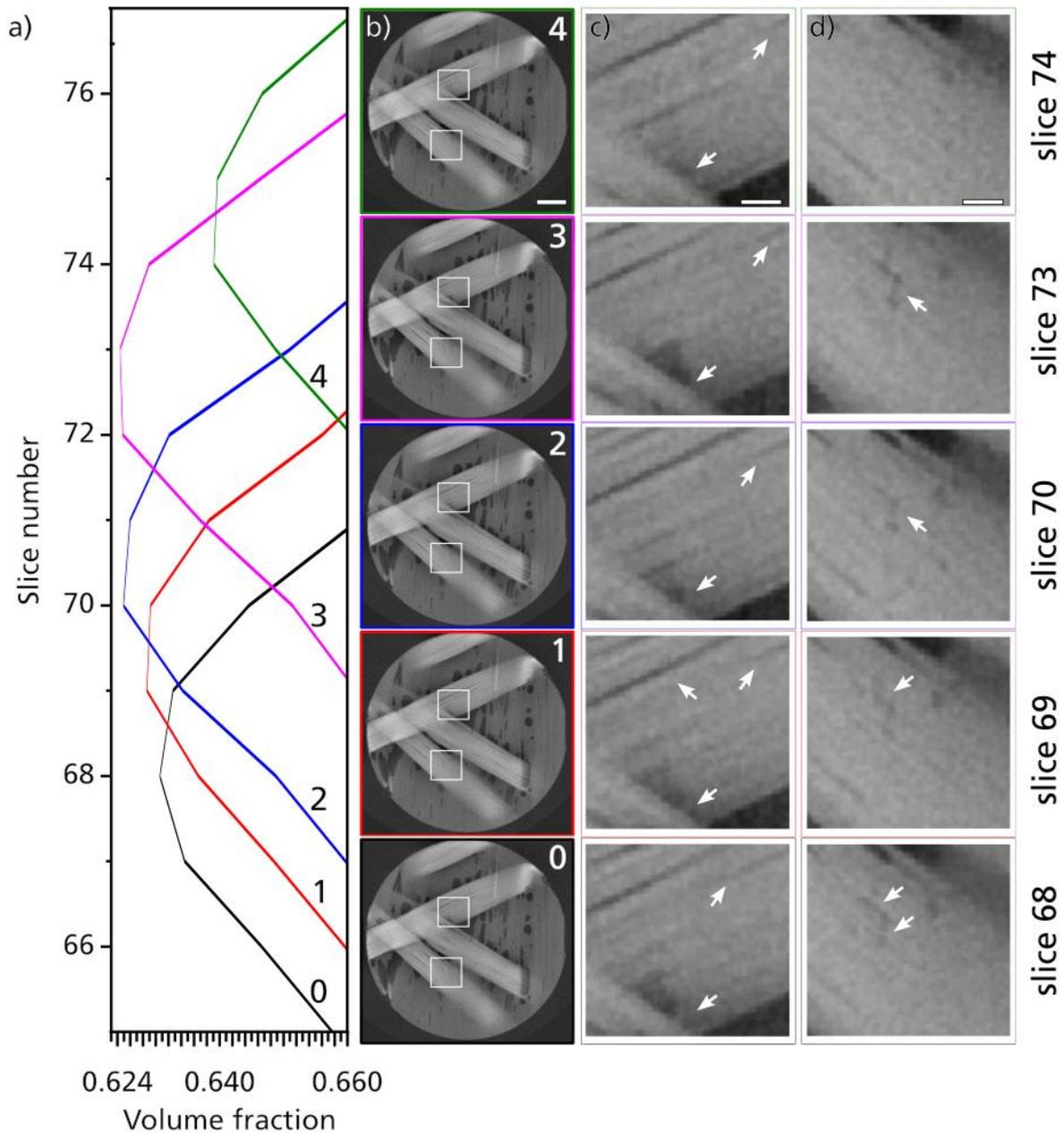

**Figure 6.** a) Volume fraction for the slice group marked by an asterisk in Fig. 5c, showing intense volume fraction change, b) tomogram images for the respective curves (scale bar: 1000 µm), with enlarged areas within a white square shown immediately on the right side (scale bar: 200 µm).

The larger displacements occurring at the bottom of the specimen are also observed by DVC displacement calculation (Fig. 7 a-d). The strain calculation performed allowed to map the strain due to the compression (Fig. 7 e-h) and shows an inhomogeneous strain distribution that appears in an angle. The strain seems to dislocate vertically with regions of small and large strain changing between compression events (Fig. 7 e-h).

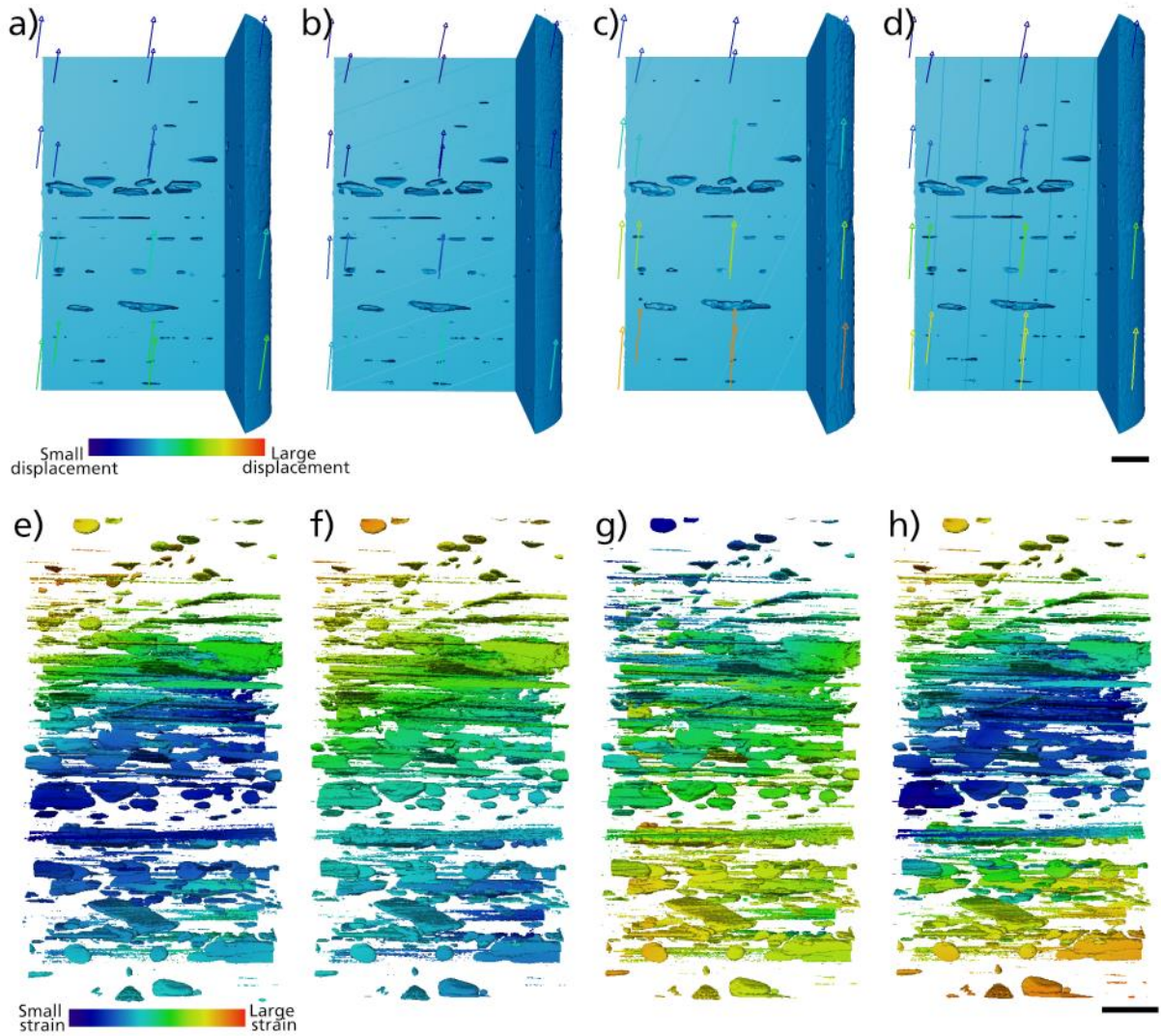

**Figure 7.** Displacement vectors computed between compressions: a) 0-1, b) 1-2, c) 2-3, and d) 3-4 superimposed onto the volumetric rendering of 0, 1, 2, and 3, respectively. Three-dimensional visualization of the axial strain in the sample under uniaxial vertical compression, with data computed between compressions: e) 0-1, f) 1-2, g) 2-3, and h) 3-4 for the same with maps shown on the sample's pores. Scale bars: 1000 μm.

**Discussion**

Here we studied the mechanical response of an unreinforced polybenzoxazine polymer and its composite reinforced with carbon fibers oriented either longitudinally or transversely to the compression axis. The nominal stress-strain curves obtained from uniaxial compression tests (Fig. 1, Table 2) show that the addition of carbon fibers in a particular orientation plays a crucial role in determining the mechanical properties of the material: when the fibers are transversely oriented with respect to the compression axis, the material is weaker than the unreinforced counterpart, while when the fibers are oriented longitudinally to the compression axis, the material supports many folds more pressure. This result is due to an interplay of factors that affect the dissipation of energy and the propagation of failure at the microstructural level in polymer composites. [22] In these materials, crack initiation is highly dependent on events such as cooperative fiber buckling, and laminae kinking. [23] Additionally, void collapse during compression contributes to the overall reduction in the stiffness of the material, and a low shear strength between matrix and fibers leads to delamination, contributing to crack propagation. [24] Our observations have shown that when the fibers were oriented transversely to the compressive load, fracture predominantly occurred along the fibers, resulting in the formation of what is referred to as fiber-tracks (see Fig. 1c). Our findings are consistent with prior observations in analogous polymer-carbon fiber composites, as documented by others. [25] These earlier studies also indicated that when fibers are longitudinally oriented to the load (perpendicular to the fracture plane), fiber failure becomes more prevalent, leading to a better material performance compared to the transverse orientation. Therefore, our results underscore the significance of fiber orientation in determining both mechanical properties and deformation characteristics.

We further studied the weaker composite specimen using CT to visualize small changes in its internal structure due to the effects of compression. CT allowed a better understanding of the fiber distribution and visualization of the defects/irregularities present in the material, mainly seen as pores in the interface of the fibers with the matrix. CT enabled imaging the wavy regions, which are known to be able to trigger local instability that leads to sudden kinking failure of the composite [26] (Fig. S4).

After an initial CT imaging, the sample was subjected to successive compression events and imaged with CT in between until approximately 4% strain to observe the effects on the specimen before extensive fiber breakage or yarn-matrix debonding, which is expected to happen by 15% strain. [27] We then used DVC to measure three-dimensional displacements and calculate strain fields by tracking changes between multiple reference volumes within the sample material in its original non-deformed state and after each deformation step to determine the corresponding displacement vectors. [28] This allows a comprehensive analysis of the internal motion and deformation occurring within the material. Analysis of the displacements and changes in volume fractions shows that the sample underwent more deformation at the bottom, specifically the surface that was in contact with the moving bottom jaw of the load stage. The strain calculations also provided additional insight into the mechanical behavior of the reinforced polymer. The results indicated that the material responded to the applied compression by generating high strain regions that are at an angle with respect to the orientation of the fibers. They are consistent with the angle of kink fractures (Fig. 1 and S4) and show the direction of the and would then illustrate its formation. The high strain regions indicate where the kink will happen due to the applied stress, causing local cracking to relieve the concentrated shear stress in those areas.

**Conclusion**

Examination of a material failure and identification of damage mechanisms is critical in the design of longer lasting materials. Our study investigated the mechanical response of an unreinforced polymer and its reinforced composite variant with carbon fibers oriented either longitudinally or transversely with respect to the compression axis. The results clearly demonstrate the significant impact of fiber orientation on the mechanical properties of the material and corroborate with other studies in this field. When the fibers are oriented transversely, the material exhibits weaker behavior compared to the unreinforced counterpart. Conversely, when the fibers are aligned longitudinally to the compression axis, the material displays significantly higher strength and can withstand much greater pressure.

The significance of this study is based on its utilization of X-ray micro-CT to capture sequential images of the material undergoing progressive compression. These images were then employed in the measurement of strain fields through DVC. We conducted CT imaging on the weaker fiber-reinforced specimen to visualize the internal structural changes caused by compression. This imaging technique allowed us to visualize the distribution of the fibers and identify defects and irregularities, particularly in the form of pores at the fiber-matrix interface. This comprehensive and innovative analysis of internal motion and deformation using DVC provided valuable information about the material's behavior at various stages of loading, revealing that the sample experienced deformation heterogeneously. The strain calculations indicated that the reinforced polymer responded to compression by generating strain parallel to the orientation of the fibers. This emphasizes the significant influence of fiber alignment on the material's mechanical response to external forces.

**Acknowledgement**

We thank Dr. Ralf Schlegel and Matthias Menzel (Fraunhofer IMWS - Halle) for the assistance with the compression test measurements and CT-imaging, respectively.

# Supplementary information

**Direct observation of the morphological changes in a carbon fiber composite by means of *in-situ* micro-CT compression experiment**

M. I. Santos[1,2], K. Koschek[3], L. Pursche[3], J. M. de Souza e Silva[2]

[1] Institute of Physics, Martin-Luther-Universität Halle-Wittenberg, Halle, Germany

[2] Fraunhofer Institute for Microstructure of Materials and Systems IMWS, Halle (Saale), Germany

[3] Fraunhofer Institute for Manufacturing Technology and Advanced Materials IFAM, Bremen, Germany,

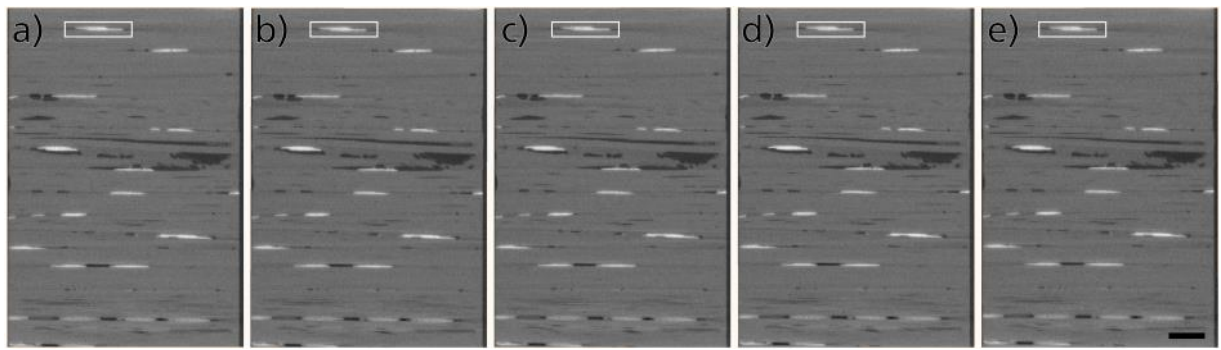

**Figure S1.** Feature used as reference for image registration shown within white box in for steps a) 0 – uncompressed to b) – e) successive compressions from 1 to 4, respectively.

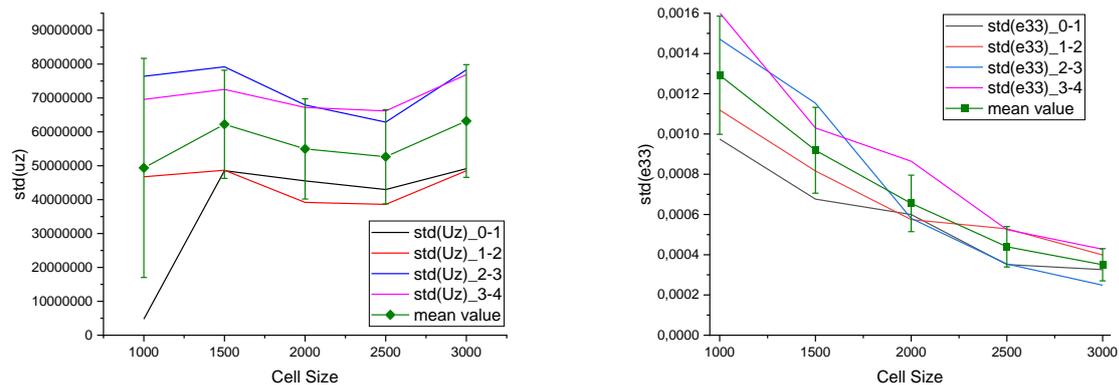

**Figure S2.** Uncertainty measurement.

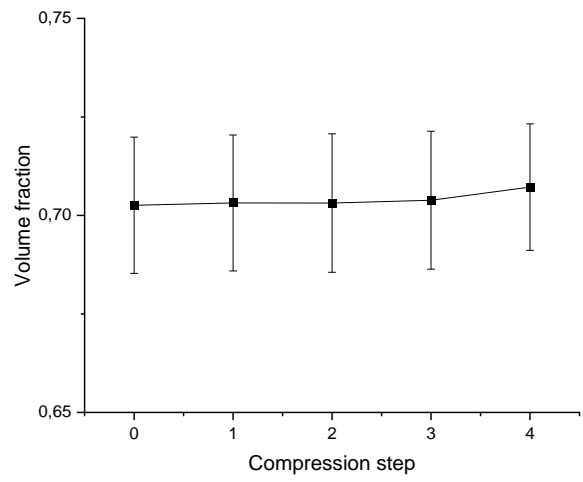

**Figure S3.** Volume fraction change for the specimen under compression.

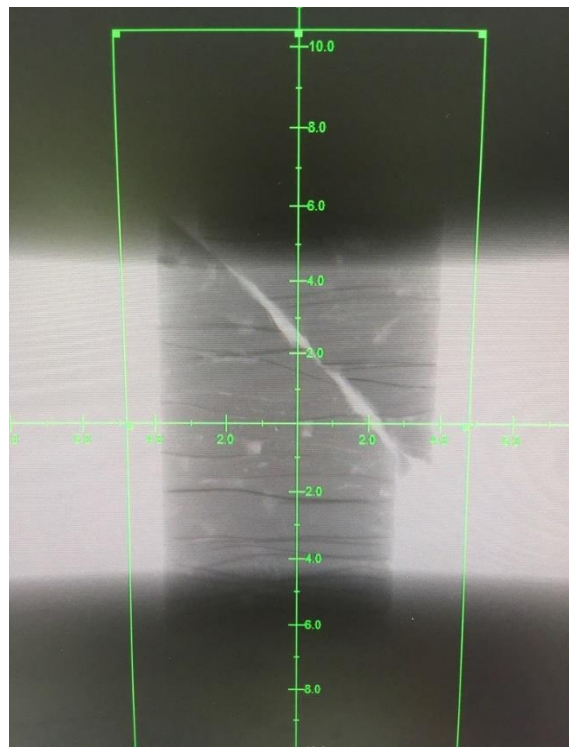

**Figure S4.** X-ray projection of the sudden kinking failure of the composite.